\documentclass[lettersize,journal]{IEEEtran}
\usepackage{amsmath,amsfonts}
\usepackage{algpseudocode}
\usepackage{algorithm}
\usepackage{array}
\usepackage{subcaption}
\usepackage{textcomp}
\usepackage{stfloats}
\usepackage{url}
\usepackage{xcolor}
\usepackage{verbatim}
\usepackage{siunitx}
\usepackage{graphicx}
\usepackage{cite}
\usepackage{enumitem}
\usepackage{multirow}
\usepackage{caption}      % often already loaded by many classes
\usepackage{bm}
\usepackage{amssymb}
 \usepackage{hyperref}
\begin{document}

\title{6G Twin: Hybrid Gaussian Radio Fields for Channel Estimation and Non-Linear Precoder Design for Radio Access Networks}

\author{Muhammad Ahmed Mohsin,~\IEEEmembership{Graduate Member, IEEE}, Muhammad Umer, Ahsan Bilal, Muhammad Ali Jameshed,~\IEEEmembership{Senior Member, IEEE},  Dean F. Hougen, John M. Cioffi,~\IEEEmembership{IEEE Fellow}
        % <-this % stops a space
\thanks{Muahmmad Ahmed Mohsin, Muhammad Umer, kand John M. Cioffi are associated with Stanford University. Email: \{muahmed, mumer, cioffi\}@stanford.edu.}% 
\thanks{Ahsan Bilal and Dean F. Hougen are associated with University of Oklahoma. Email: \{a.bilal-1, hougen\}@ou.edu.}

\thanks{Muhammad Ali Jamshed is associated with University of Glasgow. Email: muhammadali.jamshed@glasgow.ac.uk.}}

\markboth{Submitted to IEEE Transactions on Wireless Communications}%
{Shell \MakeLowercase{\textit{et al.}}: A Sample Article Using IEEEtran.cls for IEEE Journals}


\maketitle

\begin{abstract}
This work introduces \textit{6G Twin}, the \emph{first end-to-end} artificial intelligence (AI)-native radio access network (RAN) design that unifies (i) neural Gaussian Radio Fields (GRF) for compressed channel state information (CSI) acquisition, (ii) continual channel prediction with handover persistence, and (iii) an energy-optimal nonlinear precoder (\texttt{minPMAC}). GRF replaces dense pilots with a sparse Gaussian field, cutting pilot overhead by \textbf{$\sim$100$\times$} while delivering \textbf{$\sim$1.1\,ms} inference and $\leq$\textbf{2\,min} on-site training, thus enabling millisecond-scale closed-loop operation. A replay-driven continual learner sustains accuracy under mobility and cell transitions, improving channel normalized mean square error (NMSE) by \textbf{$>$10\,dB} over frozen predictors and an additional \textbf{2-5\,dB} over uniform replay, thus stabilizing performance across UMi/UMa handovers. Finally, \texttt{minPMAC} solves a convex, order-free MAC precoder design that recovers the globally optimal order from Broadcast Channel (BC) duals and minimizes transmit energy subject to minimum-rate guarantees, achieving \textbf{4–10$\times$} lower energy (scenario dependent) with \emph{monotonically increasing} bits/J as SNR grows; this translates to up to \textbf{5$\times$} higher data rate at comparable power or the same rates at substantially lower power. Together, these components form a practical, GPU-ready framework that attains real-time CSI, robust tracking in dynamic networks with efficient handovers, and state-of-the-art throughput-energy tradeoffs under 3GPP-style settings.
\end{abstract}


\section{Introduction}
Emerging 6G Networks aim for high spectral efficiency and low latency to support applications like virtual reality (VR), extended reality (XR), and internet of everything (IoE) with immersive connectivity (data rate $\geq$ 500mbps). Achieving these goals hinges on massive MIMO systems that form highly-directional beams. These in turn require fast, precise beam forming via transmitter precoding\footnote{\textbf{It is note in this paper that we assume a duality between broadcast channel and multiple access channel and use \texttt{minPMAC} to address precoder designs.}}. However, precoding critically depends on timely channel state information (CSI) at the transmitter, and the conventional pilot based CSI renders a feedback delay time of more than 8 ms causing a bottleneck to MIMO beam forming. AI for CSI prediction in~\cite{jiang2025ai} summarizes CSI overhead and channel aging issues in modern massive MIMO (mMIMO). Accurate CSI enables the transmitters to adapt their waveforms, power levels, and spatial precoding to channel conditions to achieve required data rates and link reliability. 

The mMIMO spatial multiplexing boosts data rates; with $N$ antennas at each end, the theoretical maximum achievable rates scales by a factor of $min(N_{t}, N_r, \rho_h)$ where $N_t, N_r, \rho_h$ are the number of transmitter antennas, receiver antennas and channel rank respectively. Although this increases the data rate, the pilot overhead for CSI estimation in current systems can be as high as 20\%- 40\% (on some respective tones) of each coherence interval. For instance, with a coherence time of 1 ms and a pilot duration of 0.2 ms, the overhead is 20\% and if coherence time in a dynamic environment drops to 0.5ms then the overhead increases substantially to 40\%, on that respective tone.

~\cite{fayad2024toward} documents tens of Gb/s rate and mitigates time variation. 5G~New~Radio (NR) schedules resources in units called \emph{physical resource blocks} (PRBs). 
Per 3GPP Release~18, a PRB occupies 12 contiguous sub-carriers across the full length of one slot, which, under normal cyclic prefix corresponds to 14 OFDM symbols. 
Hence, a single PRB contains 
\(
N_{\text{PRB}} = 12 \times 14 = 168
\) 
resource elements in the downlink grid. With $\mu=1$ numerology (30~kHz sub-carrier spacing), the pilot overheads are $18/168 \approx 11\%$ and $36/168 \approx 21\%$ of the transmission bandwidth.

\begin{figure}
    \centering
    \includegraphics[width=0.6 \linewidth]{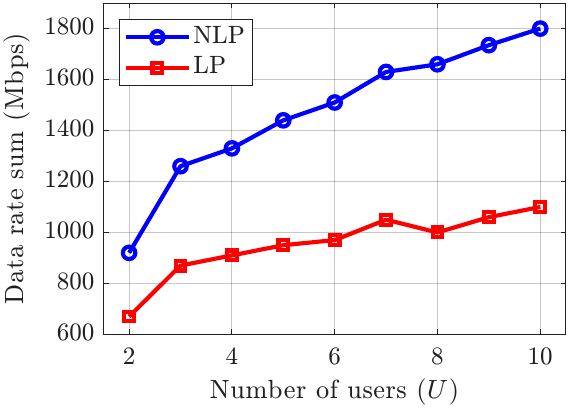}
    \caption{Linear vs. Non-Linear Precoder under rank deficient scenarios for FWA}
    \label{fig:nlp_vs_lp}
\end{figure}

Along with the CSI bottleneck, the current linear precoders do not perform well under rank-deficient scenarios, e.g, where the total number of device's receiver antennas exceeds the number of transmit antennas, as in Fig~\ref{fig:nlp_vs_lp}. Conventional orthogonal multiple-access (OMA) power-allocation schemes implemented in the latest Wi-Fi revisions—IEEE 802.11 b/g/n/ac/be—use linear reception strategies that treat the linearly processed interference from co-scheduled users as additive noise.~\cite{kerbl20233d} overviews linear MU-MIMO and OMA assumptions in practice. When the underlying MIMO channel is low rank, single-user decoding leaves the rich inter-user crosstalk unexploited, resulting in severe spectral-efficiency penalties. Moreover, meeting the gigabit-class uplink throughput targets of Wi-Fi 7 with these linear-receiver architectures necessitates transmit powers that are incompatible with the stringent energy budgets of battery-constrained hardware, e.g., augmented-reality (AR) glasses. Hence, next-generation WLANs require power-allocation and receiver designs that can jointly harness inter-user interference while respecting ultra-low-power operating limits.

State-of-the-art non-orthogonal multiple-access (NOMA) power-allocation algorithms deliver appreciable gains in spectral efficiency and energy consumption relative to their OMA counterparts~\cite{noma6, tse_viswanath_2005}. In the single-input single-output (SISO) regime, the optimal successive-interference-cancellation (SIC) schedule is trivially ordered by the users’ channel magnitudes~\cite{bhattacharya2024optimalpowerallocationtime, mohsin2025optimumpowerallocationlow}. This ordering principle, however, does not generalize to multi-user MIMO channels~\cite{wesel2002achievable, hamedoon2024towards, halabouni2025noma, fanibhare2024study}, where no inherent hierarchy emerges from the vector-valued channel realizations~\cite{noma7}. Consequently, most MIMO-NOMA studies resort to ad-hoc SIC heuristics, e.g., limiting the system to two users~\cite{10464446}, forming disjoint user pairs and applying NOMA only within each pair~\cite{noma9}, and similar simplifications. Such heuristic scheduling degrades both achievable rates and energy efficiency, falling short of contemporary WLAN requirements.

A further limitation of the prior art is its implicit assumption that all users share the entire system bandwidth, confining multiplexing to the power (or code) domain alone. Multi-carrier hybrid NOMA (MC-NOMA)~\cite{noma11} extends this framework by jointly optimizing power and sub-carrier allocation in the downlink~\cite{ghasemlou2024novel}, yet even here the SIC order is still chosen by a scalar channel-gain metric, an approach that breaks down when the access point (AP) or the users are equipped with multiple antennas. Robust, globally optimal decoding-order work for MIMO MC-NOMA appear in this problem, critical to realizing the full throughput and energy-saving potential of next-generation wireless LANs.

\textbf{Contributions. }To the best of the current knowledge, this study presents \emph{first end-to-end} AI-native RAN design that jointly covers channel acquisition (GRF), continual channel prediction with handover persistence, and energy-optimal nonlinear scheduling (\texttt{minPMAC}). (i) \textbf{GRF} (neural Gaussian Radio Fields) compress CSI acquisition, cutting pilot overhead by \textbf{$\sim$100$\times$} while delivering \textbf{$\sim$1.1\,ms} inference and state-of-the-art reconstruction accuracy across indoor/outdoor MIMO settings; this enables sub-millisecond closed-loop updates without sacrificing fidelity. (ii) \textbf{Continual learning for prediction \& handovers} uses experience replay (LARS) to adapt online under mobility and cell transitions, improving channel NMSE by \textbf{\textgreater\,10\,dB} over static predictors and a further \textbf{2–5\,dB} over uniform replay, all while maintaining accuracy across UMi/UMa handovers without per-scenario retraining. (iii) \textbf{\texttt{minPMAC}} is a convex, order-free MAC polymatroid design that recovers the globally optimal SIC sequence from dual variables and minimizes transmit energy for given minimum-rate targets; empirically it reduces the required energy by \textbf{4–10$\times$} (scenario-dependent) and yields monotonically increasing bits/J with SNR. Together, these components form a practical, GPU-ready pipeline that achieves real-time CSI, robust tracking through dynamic network conditions and handovers, and substantial energy savings under multiuser MIMO operation.

\begin{table*}[t]
\caption{Key notation used in the paper.}
\label{tab:notation}
\centering
\footnotesize
\renewcommand{\arraystretch}{1.15}
\begin{tabular}{p{3.2cm} p{11cm}}
\hline
\textbf{Symbol} & \textbf{Description} \\
\hline
$U$ & Number of users; index set $\mathcal U=\{1,\dots,U\}$. \\
$N$ & Number of OFDM subcarriers (tones); index set $\mathcal N=\{1,\dots,N\}$. \\
$L_{x,u}$, $L_y$ & Transmit antennas of user $u$; receive antennas at BS. \\
$\mathbf H_{u,n}\!\in\!\mathbb C^{L_y\times L_{x,u}}$ & MIMO channel matrix of user $u$ on tone $n$. \\
$\widehat{\mathbf H}_{t},\ \widehat{\mathbf H}_{t+\Delta}$ & GRF channel estimate at time $t$; $\Delta$-step forecast from the predictor. \\
$\mathbf R_{xx}(u,n)\succeq\mathbf 0$ & Transmit covariance of user $u$ on tone $n$. \\
$\mathbf R_{nn}=\sigma^2\mathbf I_{L_y}$ & Receiver noise covariance. \\
$b_{u,n}$,\; $\mathbf b_{\min}$ & Achievable rate of user $u$ on tone $n$ (bits/s/Hz); per-user minimum-rate targets. \\
$w_u\!\ge\!0$ & Energy weight for user $u$ in the precoder objective. \\
$\mathcal T\subseteq\mathcal U$ & User subset used in MAC polymatroid constraints. \\
$\mathbf X_t$ & Sliding CSI window of length $T$ used by the predictor. \\
$\mathcal M,\ N_{\mathrm{buf}}$ & Replay buffer and its size for continual learning. \\
$\lambda\!\in\![0,1]$ & Mixture weight for current vs.\ replay loss in training. \\
$\mathrm{NMSE}$ & Normalized MSE: $\|\mathbf H-\widehat{\mathbf H}\|_F^2/\|\mathbf H\|_F^2$. \\
$\mathrm{SNR}_{\mathrm{dB}}$ & $10\log_{10}\!\left(\|\mathbf H_{\text{gt}}\|_F^2/\|\mathbf H_{\text{pred}}-\mathbf H_{\text{gt}}\|_F^2\right)$. \\
GRF & Gaussian Radio Fields (compressed CSI acquisition). \\
LARS & Loss-Aware Reservoir Sampling (replay buffer management). \\
\texttt{minPMAC} & Energy-optimal nonlinear precoder on the MAC polymatroid. \\
\hline
\multicolumn{2}{p{14.6cm}}{%
 Boldface upper-case (lower-case) letters denote matrices (vectors); 
$\operatorname{tr}(\cdot)$ is the trace, $(\cdot)^{\mathrm H}$ the Hermitian transpose, 
and $\|\cdot\|_{\mathrm F}$ the Frobenius norm. 
The set $\{1,\dots,N\}$ indexes OFDM sub-carriers, and time is slotted at pilot instants $t\in\mathbb N$. 
For any quantity $A$, the notation $A_{t}$ (resp.\ $A_{t+\Delta}$) refers to the present (resp.\ predicted) slot.} \\
\hline
\end{tabular}
\end{table*}

\section{Related Work}
\textbf{Channel Estimation.} Neural radiance-field (NeRF) approaches such as NeRF$^2$~\cite{Zhao_2023} and NeWRF~\cite{lu2024newrf} give much better channel-estimation accuracy than classic methods. Still, they need very dense measurements (about 178 samples per cubic foot), hours of GPU training, and 200–350 ms inference times, far too slow for real-time vehicle links. To speed inference time, researchers turned to 3D Gaussian Splatting (3DGS): WRF-GS~\cite{wen2025neural} uses spherical-harmonic features on 3DGS to map RF power, but this ignores full MIMO CSI and depends on a Mercator projection that does not match electromagnetic physics; RadSplatter~\cite{niemeyer2025radsplat} improves RSSI prediction with a relaxed-mean features yet suffers unstable optimization. RFSPM~\cite{yang2025scalable} gains speed by adding tiny multi-layer perceptrons (MLPs) per Gaussian and custom CUDA kernels, making deployment harder. In short, both NeRF and 3DGS based models borrow optical-imaging assumptions that fail to capture true 3-D radio wave superposition and still demand too much data or compute for millisecond-level wireless use~\cite{yang2025generative, 8929013}.

\textbf{Channel Prediction.} Data-driven channel prediction has evolved from  long-short term memory (LSTM) model baselines to transformers, large language models, and diffusion networks. Recent studies use transformers for the channel prediction~\cite{zhang2024transformer}. LSTMs outperform autoregressive integrated moving average (ARIMA) but stumble on long memories~\cite{article,jiang2020recurrent}; transformers ease that issue yet demand heavy, matched condition training and large computation, and their error jumps 15–30\% when array spacing changes~\cite{jiang2022accurate,zhang2024transformer,kim2025machine}. Generative pretrained transformer (GPT-2) adaptation enables near-zero-shot multiple input single output orthogonal frequency division multiplexing (MISO-OFD)M prediction after brief fine-tuning, but the language-to-CSI gap still caps accuracy~\cite{liu2024llm4cp}. Diffusion models can synthesize scarce MIMO data or handle joint source-channel estimation~\cite{lee2024generatinghighdimensionaluserspecific,bhattacharya2025sic,zilberstein2024joint, hu2024transfer}, yet they remain slow and lack online drift adaptation. These limitations motivate a continual-learning MIMO predictor. 

To break this accuracy–complexity log-jam, each link is bootstrapped with a few high-fidelity snapshots from a NeRF-based estimator, then a lightweight continual-learning predictor  updates online as the channel evolves. This single pipeline keeps the instant CSI precision of NeRF during the initial iterations while avoiding its data hunger and heavy retraining thereafter, yielding a unified framework for scalable MIMO estimation and long-horizon prediction under changing propagation conditions.

\textbf{NOMA Methods} Power‐domain NOMA is widely recognized to outperform OMA at a given power budget—especially in SISO settings where SIC follows the users’ channel strengths—and extensive surveys establish these fundamentals and their 5G/6G relevance~\cite{Ding2017PIEEE,Islam2019Overview}. However, once channels are vector‐valued (MIMO), there is no natural scalar ordering; most MIMO–NOMA designs therefore rely on \emph{heuristics} such as two‐user pairing per beam/subcarrier, fixed or gain‐based SIC schedules, or small disjoint clusters—choices that simplify scheduling but break global optimality and often induce non-convex formulations~\cite{Ding2017PIEEE}. MC-NOMA improves matters by co-optimizing power and subcarriers, yet mainstream formulations still impose two-users-per-subchannel and gain-based SIC rules, so the core limitations persist~\cite{Di2016MCNOMA}. In contrast, \emph{\texttt{minPMAC}} treats the capacity region explicitly as a polymatroid, optimizes power–subcarrier \emph{covariances} without embedding the decoding order (thus preserving convexity), and then \emph{recovers} the globally optimal SIC sequence from the dual structure—thereby avoiding combinatorial search; the resulting MAC solution maps cleanly to the downlink via classical uplink–downlink duality~\cite{TseViswanath2005,Hanly1998Polymatroid,Viswanath2003Duality,Weingarten2006, khan2025machinelearningdrivenperformanceanalysis, Cioffi_GDFE_Ch5, Cioffi1997}.

\section{Problem Scope and System Architecture}
A single‐cell multi–carrier MIMO uplink model has
$U$ users, each equipped with $L_{x,u}$ antennas. These transmit to a single radio twin base station (BS) equipped with $L_{y}$ receive antennas. The BS hosts the proposed \emph{AI–driven digital twin} composed of three sequential modules. First, \emph{Gaussian Radio Fields} (GRF), performs compressed CSI acquisition by rendering the instantaneous MIMO channel from a learned set of 3‐D Gaussian primitives. Second, a continual‐learning (CL) channel predictor exploits temporal correlations to forecast future CSI and thereby compensates for channel aging as well as mobility‐induced hand-over drift~\cite{mohsin2025continuallearningwirelesschannel}. Third, the \emph{\texttt{minPMAC}} nonlinear precoder consumes the predicted channel and outputs the globally optimal power–sub-carrier allocation together with the SIC decoding order. The outputs include optional time‐sharing coefficients when multiple orders are required~\cite{bhattacharya2025optimum}. All three stages execute on the BS's GPU cluster, enabling a sub-millisecond closed-loop adaptation cycle. The next sections discuss theoretical and experimental details of each module separately.\\
\textbf{Channel Estimation Formulation. }Let $\mathbf E(\mathbf r,\omega)\!\in\!\mathbb C^{3}$ denote the steady‐state electric field at spatial
location $\mathbf r\!\in\!\mathbb R^{3}$ and angular frequency $\omega$. In a linear, source‐driven medium,
the field obeys the vector Helmholtz equation
\begin{equation}
\label{eq:helmholtz}
\nabla\times\nabla\times\mathbf E(\mathbf r,\omega)
      -k^{2}(\omega)\,\mathbf E(\mathbf r,\omega)
      \;=\;
      -j\omega\mu_{0}\,\mathbf J(\mathbf r,\omega),
\end{equation}
where $k(\omega)=\omega\sqrt{\mu\epsilon}$ is the wave number and
$\mathbf J$ the impressed current density. The \emph{channel state
information} (CSI) between a transmit array with phase centers
$\{\mathbf r^{\mathrm{tx}}_{m}\}_{m=1}^{L_{x}}$ and a reception array
$\{\mathbf r^{\mathrm{rx}}_{\ell}\}_{\ell=1}^{L_{y}}$ is then
captured by the frequency‐dependent MIMO transfer matrix
\begin{equation}
\label{eq:channel-matrix}
\bigl[\mathbf H(\omega)\bigr]_{\ell,m}
   \;=\;
   \mathbf g^{\!\top}\!
   \mathbf E\!\bigl(\mathbf r^{\mathrm{rx}}_{\ell},\omega;
                     \mathbf r^{\mathrm{tx}}_{m}\bigr),
\end{equation}
ith $\mathbf g$ the receiver’s effective dipole vector.
Obtaining \eqref{eq:channel-matrix} from sparse pilot measurements is
an ill‐posed \emph{inverse electromagnetic problem} that must infer
the volumetric Green’s function satisfing \eqref{eq:helmholtz}
under highly heterogeneous, partly unknown, boundary conditions.
\begin{figure*}[t!]
    \centering
    \includegraphics[width=\linewidth]{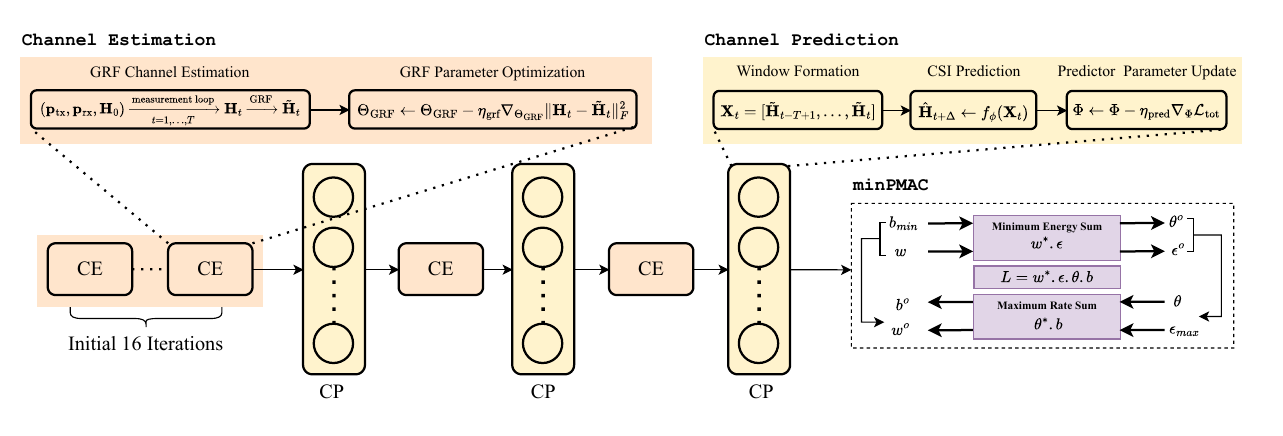}
    \caption{Flow diagram of the proposed hybrid RF-channel pipeline.
During the first 16 iterations, a NeRF-based Gaussian Radiance Field (GRF) module performs high-fidelity channel estimation (CE). Once initial parameters converge, the system transitions to a lightweight continual-learning predictor (CP) that: (i) forms sliding CSI windows, (ii) updates model weights online, and (iii) delivers long-horizon forecasts alongside link-quality metrics—minimum data-rate (\texttt{minPMAC}), optimal decoding order, receiver matrix, and minimum energy-rate.}
    \label{fig:flowchart}
\end{figure*}
A direct numerical solution—e.g., via finite‐difference time‐domain
(FDTD) or finite‐element methods—samples the domain at
sub‐wavelength resolution, leading to linear systems of order
$\mathcal O\!\bigl((\ell/\lambda)^{3}\bigr)$ for an environment of
characteristic size $\ell$. Such solvers incur terabyte‐scale memory
footprints and runtimes far beyond the milliseconds coherence interval
of 5G/6G links. Geometric optics (ray‐tracing) offers an
$\mathcal O(N_{\text{ray}}N_{s}N_{b}^{R})$, where $N_{ray}$ is the number of rays, $N_s$ is the number of surfaces and $N_b^R$ is the number of depth rays, surrogate by taking the $\lambda\!\to\!0$ limit, but these neglect diffraction and scattering, and still become prohibitive once the number of surfaces $N_{s}$ and bounce depth $R$ grow in realistic indoor or urban scenes. Consequently, state‐of‐the‐art CSI estimation adopts data‐driven
approximations that trade physical fidelity for tractability in
compressive pilot recovery, learned NeRF‐style implicit fields, or,
as proposed here, the explicit \emph{Gaussian Radio Field}
representation. The latter replaces the volumetric mesh by a sparse
set of adaptive anisotropic Gaussian primitives, yielding closed‐form
superposition in \eqref{eq:channel-matrix} and enabling sub‐millisecond
channel synthesis suitable for real‐time digital‐twin operation.\\

\textbf{minPMAC Precoder Formulation. }Let $\mathcal U\!=\!\{1,\dots,U\}$ index the users and
$\mathcal N\!=\!\{1,\dots,N\}$ the OFDM sub–carriers.
For user $u\!\in\!\mathcal U$ on tone $n\!\in\!\mathcal N$, $\mathbf R_{xx}(u,n)\!\in\!\mathbb C^{L_{x,u}\times L_{x,u}}$ is the
transmit covariance (equivalently, power‐allocation) matrix and $b_{u,n}$ is the achievable rate under a fixed but \emph{unspecified}
SIC decoding order. The overall design objective is to meet a vector rate requirement $\mathbf b_{\min}\!=\![b_{\min,1},\dots,b_{\min,U}]^{\!\top}$
while minimizing a weighted sum of user energies.
Formally, the task is cast as the following
\emph{energy–sum minimization} problem:
\begin{equation} \label{eq:primal-energy}
\begin{aligned}
&\min_{\left\{R_{\boldsymbol{x} \boldsymbol{x}}(u, n)\right\}} 
\quad \sum_{u=1}^U \sum_{n=1}^{N} 
w_u \cdot \operatorname{trace}\!\left\{R_{\boldsymbol{x} \boldsymbol{x}}(u, n)\right\} \\
&\text{C1}: \mathbf{b}=\sum_{n=1}^{N}\!\left[b_{1, n}, b_{2, n}, \ldots ,b_{U, n}\right]^T 
\succeq \mathbf{b}_{\min } \succeq \mathbf{0} \\
&\text{C2}: \sum_{u \in T}\sum_{n=1}^{N} b_{u, n} \;\leq\;
   \log_2 \Biggl| R_{\boldsymbol{nn}} 
   + \sum_{u\in T} H_{u, n} R_{\boldsymbol{x}\boldsymbol{x}}(u, n) H_{u, n}^* \Biggr| \\
&\qquad\qquad\qquad - \log_2 \bigl| R_{\boldsymbol{nn}} \bigr|,
\qquad \forall T \subseteq \{1, 2, \dots, U\} \\
&\text{C3}: \mathbf{R}_{\boldsymbol{x}\boldsymbol{x}}(u,n) \succeq \mathbf{0}, 
\quad \forall u \in \{1, 2, \dots, U\}, \;
\forall n \in \{1, 2, \dots, N\}
\end{aligned}
\end{equation}
where $\mathbf H_{u,n}\!\in\!\mathbb C^{L_{y}\times L_{x,u}}$ is the MIMO
channel for user~$u$ on tone~$n$,
$\mathbf R_{nn}\!=\!\sigma^{2}\mathbf I_{L_{y}}$ \footnote{Note here that $H_{u,n} = R_{nn}^{-1/2}H$ which allows use of an equivalent white noise model.} is the noise covariance
matrix, and
$w_u\!\ge\!0$ assigns relative importance to user~$u$’s transmit
energy. 
Constraint \textbf{C1} enforces the per–user minimum–rate targets,
while \textbf{C2} guarantees that \emph{every} subset of users
$\mathcal T$ operates inside the MAC capacity polymatroid.
Finally, \textbf{C3}
ensures that each covariance matrix is Hermitian positive semi‐definite. Problem~\eqref{eq:primal-energy} is a
convex semidefinite program:
the objective is linear in the decision variables
$\mathbf R_{xx}(u,n)$, 
\textbf{C1} is affine,  
\textbf{C2} is jointly concave in $\{\mathbf R_{xx}(u,n)\}$ by MIMO
capacity theory,  
and \textbf{C3} defines a convex cone.
Crucially, the formulation is \emph{decoupled} from the SIC decoding
order~$\boldsymbol\pi$; therefore no non-convex `ordering variables’
are introduced, in contrast to heuristic approaches
that embed the order into the optimization
and render the problem non-convex
\cite{noma5,noma6,noma7,noma8}. Associating dual variables $\{\theta_u\}$ with the minimum‐rate
constraints~\textbf{C1} yields the Lagrange dual
\emph{weighted sum–rate maximization under an energy budget},
from which the globally optimal decoding order can be inferred
(\S\ref{subsec:decoding-order}).
Thus, the overall design procedure proceeds in two convex steps:
(i) solve~\eqref{eq:primal-energy} to obtain the
power–sub-carrier allocation, and
(ii) determine the SIC sequence by sorting the optimal dual
multipliers. This separation preserves convexity and ensures an
efficient, globally optimal solution pipeline for nonlinear precoding. The resultant $R_{xx} (u,n)$, along with learned $H_{u,n}$, then must accordingly be used in a GDFE that the \texttt{minPMAC} optimization assumes.

\begin{figure}
    \centering
    \includegraphics[width=\linewidth]{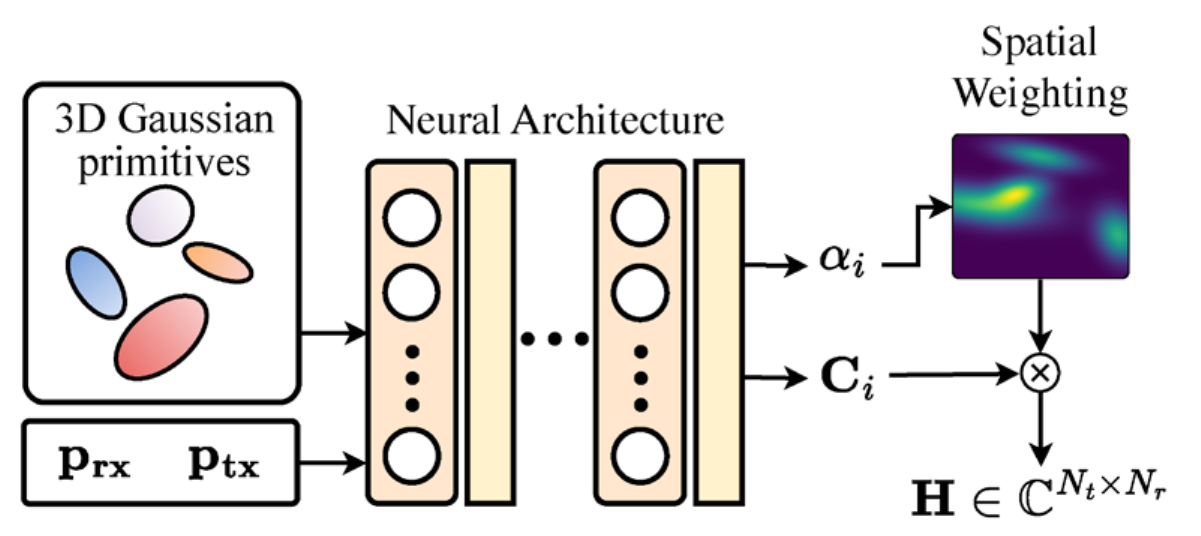}
  \caption{
    Schematic of the differentiable MIMO channel model. Transmitter and receiver positions $\mathbf{p}_{\mathrm{tx}}$ and $\mathbf{p}_{\mathrm{rx}}$ are fused with $M$ learnable 3D Gaussian primitives (means $\boldsymbol\mu_i$, covariances $\boldsymbol\Sigma_i$) and passed through a neural network to produce, for each primitive $i$, a base activation $\alpha_i$ and a complex channel contribution $\mathbf{C}_i$. Each contribution is gated by the spatial weight 
    $w_i(\mathbf{p}_{\mathrm{rx}},\boldsymbol\mu_i,\boldsymbol\Sigma_i)$ 
    and summed to yield the predicted channel matrix 
    $\mathbf{H}=\sum_{i=1}^M w_i\,\alpha_i\,\mathbf{C}_i\in\mathbb{C}^{N_t\times N_r}$. 
  } 
  \label{fig:ngrf}
\end{figure}
\section{Hybrid Gaussian Observation for Channel Estimation}
The proposed GRF model~\cite{umer2025neuralgaussianradiofields} consists mainly of 3 submodules, which are explained step-by-step as:\\
\textbf{3D Gaussian Primitive Initialization. }The proposed \emph{Gaussian Radio Field} (GRF) models the
propagation environment by a sparse set
$\mathcal G=\{\mathcal G_i\}_{i=1}^{N}$ of anisotropic
three-dimensional Gaussians. 
Primitive $i$ is characterized by its centre
$\boldsymbol{\mu}_{i}\!\in\!\mathbb R^{3}$ and
positive-definite covariance
$\boldsymbol{\Sigma}_{i}\!\in\!\mathbb S_{+}^{3}$; additional
electromagnetic attributes are supplied by the neural field
(see Fig.~\ref{fig:ngrf}). Owing to the linearity of Maxwell’s equations in passive,
non-magnetic media, the total field superposes of the
fields generated by these primitives. Each Gaussian, therefore, acts
as an adaptive basis function, chosen to capture fine spatial
channel variations.

For a query location $\mathbf x\!\in\!\mathbb R^{3}$ the unnormalized
kernel weight of primitive $i$ is
\begin{equation}
w_{i}(\mathbf x)=
\exp\!\Bigl(
      -\tfrac12(\mathbf x-\boldsymbol\mu_{i})^{\!\top}
      \boldsymbol\Sigma_{i}^{-1}
      (\mathbf x-\boldsymbol\mu_{i})
     \Bigr),
\label{eq:grf_weight}
\end{equation}
which omits the constant Gaussian prefactor because only relative
Weights are required.

Factoring $\boldsymbol\Sigma_{i}=\mathbf R_{i}\mathbf S_{i}^{2}\mathbf R_{i}^{\!\top}$, guarantees $\boldsymbol\Sigma_{i}\succ 0$ while keeping the parameters unconstrained.
Here $\mathbf R_{i}\!\in\!\mathrm{SO}(3)$ is unitary, and
$\mathbf S_{i}=\operatorname{diag}(\mathrm e^{s_{i,1}},
                                   \mathrm e^{s_{i,2}},
                                   \mathrm e^{s_{i,3}})$
ensures strictly positive eigenvalues through the exponential map.
This compact parameterization facilitates the GRF's end-to-end differentiable
optimization. \\
\textbf{Attribute Network. }The electromagnetic response of each Gaussian primitive is generated in real-time by two lightweight neural modules (Fig.~\ref{fig:ngrf}). 
The \emph{attribute network}
\(
\mathcal F_{\mathrm{attr}}:\mathbb R^{3}\!\times\!\mathbb R^{3}\!\to\!
\mathbb R^{d}\!\times\!\mathbb R
\)
receives the primitive center $\boldsymbol\mu_{i}$ and the transmitter
location $\mathbf p_{\mathrm{tx}}$, and returns a latent
descriptor~$\mathbf z_{i}\!\in\!\mathbb R^{d}$ together with a scalar
activation~$\alpha_{i}$:
\[
(\mathbf z_{i},\alpha_{i})
= \mathcal F_{\mathrm{attr}}\bigl(
     \gamma_{L}(\boldsymbol\mu_{i}),
     \gamma_{L}(\mathbf p_{\mathrm{tx}})\,;\,
     \Theta_{\mathrm{attr}}
   \bigr).
\]

where $\gamma_{L}(\cdot)$ is the multi-resolution positional encoding \cite{mildenhall2020nerf, fan2025generative} and $\mathbf{z_i}$ is the extracted features in latent space later utilized in the channel contribution $\mathbf{C_i}$, 
\begin{align}
\gamma_{L}(\mathbf x) &=
\bigl[
   \mathbf x,\,
   \sin(2^{0}\pi\mathbf x),\, \cos(2^{0}\pi\mathbf x), \dots, \nonumber\\
 &\quad
   \sin(2^{L-1}\pi\mathbf x),\, \cos(2^{L-1}\pi\mathbf x)
\bigr].
\end{align}
Choosing $L$ such that $2^{L-1}\!\ge\!\ell/\lambda$ (with
$\ell$ the scene diameter and $\lambda$ the carrier wavelength)
guarantees Nyquist coverage of spatial harmonics. Throughout, $L\!=\!16$ is sufficient for $\ell\!\approx\!50$\,m and $\lambda\!\approx\!5$\,cm.

The latent vector $\mathbf z_{i}$ captures local scattering traits,
while $\alpha_{i}$ modulates their overall strength.
A separate \emph{decoder network}
\(
\mathcal F_{\mathrm{dec}}:\mathbb R^{d}\!\to\!\mathbb R^{N_t\times N_r}\times
\mathbb R^{N_t\times N_r}
\)
maps~$\mathbf z_{i}$ to the real and imaginary parts of the complex
contribution,
\(
\mathbf C_{i}= \mathbf C^{\mathrm{re}}_{i}
             + j\,\mathbf C^{\mathrm{im}}_{i}
             =\mathcal F_{\mathrm{dec}}(\mathbf z_{i};\Theta_{\mathrm{dec}}),
\)
thereby endowing each Gaussian with an effective $N_t\!\times\!N_r$
MIMO transfer matrix.

\begin{figure*}
    \centering
    \includegraphics[width=\linewidth]{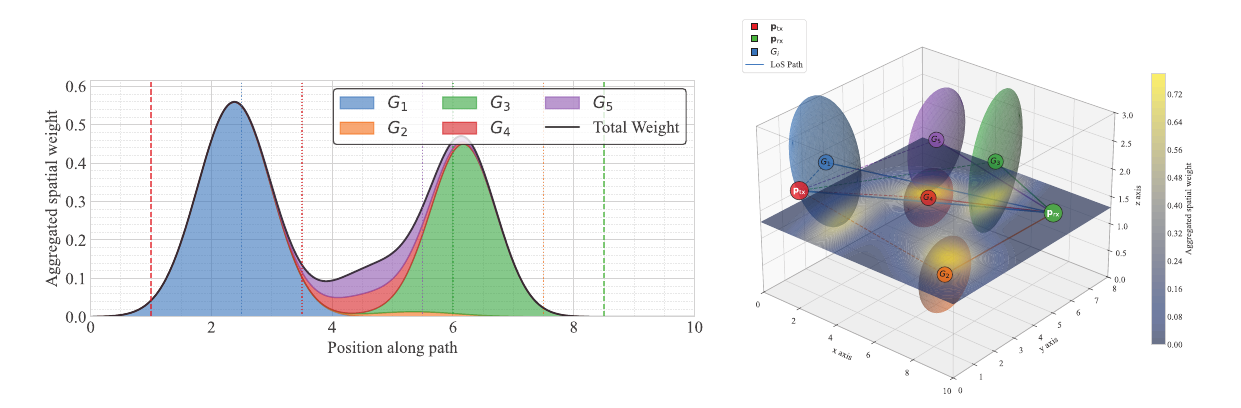}
    \caption{\textbf{Visualization of the spatial weighting in GRF.} (Right:) Anisotropic 3D Gaussians ($G_1$-$G_5$) are shown as ellipsoids with their shapes determined by covariance $\boldsymbol{\Sigma}_i$, along with the transmitter ($\mathbf{p}_{\text{tx}}$), receiver ($\mathbf{p}_{\text{rx}}$), and line-of-sight path. The heat map represents the spatial weight distribution at receiver height. (Left:) Weight profile along the $\mathbf{p}_{\text{tx}}$--$\mathbf{p}_{\text{rx}}$ path showing individual Gaussian contributions and their cumulative effect (black outline).}
    \label{fig:spatial_weights}
\end{figure*}

\textbf{Channel Rendering. }GRF evaluates the MIMO channel by a single weighted sum over
$N$ Gaussian primitives, thus bypassing ray-marching or volumetric integration. 
For a transmitter at $\mathbf p_{\mathrm{tx}}$ and a receiver at
$\mathbf p_{\mathrm{rx}}$ the $N_{t}\!\times\!N_{r}$ channel matrix is

\begin{equation}
\mathbf H(\mathbf p_{\mathrm{rx}},\mathbf p_{\mathrm{tx}})
   =\sum_{i=1}^{N}
     w_{i}(\mathbf p_{\mathrm{rx}})\,\mathbf C_{i},
\label{eq:grf_render}
\end{equation}

\noindent
where $\mathbf C_{i}$ is the complex contribution of primitive~$i$ and
the spatial weight. Figure~\ref{fig:spatial_weights} illustrates this spatial weighting mechanism.

\begin{equation}
w_{i}(\mathbf p_{\mathrm{rx}})
   =\alpha_{i}\,
     \exp\!\Bigl(
           -\tfrac12
           (\mathbf p_{\mathrm{rx}}-\boldsymbol\mu_{i})^{\!\top}
           \boldsymbol\Sigma_{i}^{-1}
           (\mathbf p_{\mathrm{rx}}-\boldsymbol\mu_{i})
         \Bigr)
\label{eq:grf_weight_full}
\end{equation}

\noindent
decays with the Mahalanobis distance between the receiver and the
Gaussian centre. 
The scalar $\alpha_{i}$ modulates the overall strength of the primitive~$i$.

Equation~\eqref{eq:grf_render} requires
$\mathcal O\!\bigl(N\,N_{t}\,N_{r}\bigr)$ operations, exploiting the
intrinsic sparsity of radio propagation; traditional NeRF/3DGS
pipelines incur
$\mathcal O(N_{\text{ray}}N_{\text{sample}}|\Theta|)$ evaluations of a
large MLP and are therefore two–to–three orders of magnitude slower.
Because GRF directly computes the field at the true 3-D receiver
location, it avoids the projection artifacts that arise when 3-D
Gaussian are rasterized onto 2-D planes. Fig.~\ref{fig:ngrf} demonstrates the attribute network and the spatial weighting pipeline of our GRF model.
\begin{algorithm}[h]
\caption{GRF Channel Rendering (\,$\mathcal O(N\,N_t\,N_r)$\,)}
\label{alg:grf_render}
\begin{algorithmic}[1]
\Require 
  primitive set $\{(\boldsymbol\mu_i,\mathbf q_i,\mathbf s_i,\alpha_i,\mathbf z_i)\}_{i=1}^{N}$,\\
  transmitter position $\mathbf p_{\mathrm{tx}}$, receiver position $\mathbf p_{\mathrm{rx}}$
\Ensure 
  channel matrix $\mathbf H(\mathbf p_{\mathrm{rx}},\mathbf p_{\mathrm{tx}})\in\mathbb C^{N_t\times N_r}$
\vspace{2pt}
\State $\mathbf H \gets \mathbf 0$ \Comment{initialize accumulator}
\For{$i=1$ to $N$}
    \State $\mathbf R_i \gets \textsc{QuatToRot}(\mathbf q_i)$
    \State $\boldsymbol\Lambda_i^{-1} \gets 
           \mathbf R_i \,\mathrm{diag}\!\bigl(e^{-2\mathbf s_i}\bigr)\,
           \mathbf R_i^{\mathsf T}$
    \State $\Delta\mathbf p \gets \mathbf p_{\mathrm{rx}} - \boldsymbol\mu_i$
    \State $w_i \gets \alpha_i \exp\!\bigl(-\tfrac12
           \Delta\mathbf p^{\mathsf T}\boldsymbol\Lambda_i^{-1}\Delta\mathbf p
           \bigr)$
    \State $\mathbf C_i \gets \textsc{Decode}(\mathbf z_i)$
    \State $\mathbf H \gets \mathbf H + w_i\,\mathbf C_i$
\EndFor \\
\Return $\mathbf H$
\end{algorithmic}
\end{algorithm}

\section{Continual Channel Tracking and Prediction} 
\begin{figure}
    \centering
    \includegraphics[width=\linewidth]{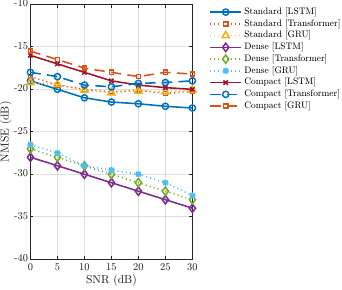}
    \caption{Baseline SNR trained on UMi dense and tested on all
scenarios under different architectures.}
    \label{fig:baseline_cp}
\end{figure}

Between two successive GRF estimates, the propagation medium may evolve appreciably (mobility, blockage, antenna reconfiguration, cell handoff), rendering a static predictor obsolete. This work, therefore, develops a continual-learning (CL) channel-prediction algorithm that adapts online across heterogeneous network configurations and handoff situations without requiring per-scenario retraining. Figure~\ref{fig:baseline_cp} shows why this is needed: offline models trained once and frozen fail to track shifting channel statistics as the UE moves from UMi‑compact $\rightarrow$ UMi‑dense $\rightarrow$ UMi‑standard (and as SNR improves). Dense variants eventually reach $-30$\,dB NMSE, but Standard and Compact topologies saturate in the $-22$ to $-25$\,dB range. This difference between the NMSE for Dense and other scenarios (Standard and Compact) gives evidence of distribution mismatch and weight staleness. The proposed CL approach interleaves each new mini‑batch with a bounded replay buffer of historical samples, regularizing updates and mitigating catastrophic forgetting as conditions drift across cells. At every GRF refresh, prediction errors are logged; high‑loss (“hard”) fades are preferentially retained via experience replay (ER)s, ensuring that rare yet performance‑critical states remain represented in the buffer. As shown in Fig.~\ref{fig:continual_cp}, this ER loop yields a step‑change in accuracy: all three topology families collapse to a low‑error band near $-38$\,dB NMSE by $\sim\!10$\,dB SNR and continue improving to $\approx -42$\,dB at $30$\,dB SNR (best: Compact‑LARS), more than 10\,dB better than the non‑CL Standard/Compact baselines and a clear gain even over Dense.

\subsection{Experience Replay for GRF-aware Prediction}
Let $\{\mathcal{D}^{(m)}\}_{m\ge 1}$ be the sequence of mini-batches collected in successive cells.
At iteration $k$, the predictor $f_{\boldsymbol{\Phi}}$ trains on the current batch $\mathcal{D}^{(k)}$ while regularizing against forgetting using experience replay drawn from $\bigcup_{m<k}\mathcal{D}^{(m)}$.
After each GRF rendering, the latest estimate $\widehat{\mathbf{H}}_{t}\in\mathbb{C}^{N_t\times N_r}$ is appended to a sliding window $\mathbf{X}_{t}=[\,\widehat{\mathbf{H}}_{t-T+1},\dots,\widehat{\mathbf{H}}_{t}\,]$, which the predictor maps to a $\Delta$-step forecast $\widehat{\mathbf{H}}_{t+\Delta}=f_{\boldsymbol{\Phi}}(\mathbf{X}_{t})$.
The forecast is consumed by the scheduler immediately, while the pair $(\mathbf{X}_{t},\widehat{\mathbf{H}}_{t+\Delta})$ is pushed to the replay buffer $\mathcal{M}$.
When the next GRF estimate $\widehat{\mathbf{H}}_{t+\Delta}$ arrives, the prediction error is logged as a replay score, ensuring that difficult samples are revisited.
Thus, GRF and the predictor form a closed loop: GRF supplies ground-truth snapshots; the predictor fills the gaps; and its continually refined outputs feed back into the digital-twin pipeline, promoting ergodic tracking of the evolving channel distribution.\\
\textbf{Replay Buffer. }A fixed-size buffer
$\mathcal M=\{e_{1},\dots,e_{N_{\text{buf}}}\}$ stores past channel
samples, where an element
$e=(\widehat{\mathbf H}_{t},\mathbf p_{\mathrm{tx}},\mathbf p_{\mathrm{rx}})$
comprises the latest GRF estimate and the corresponding link geometry. Upon observing a new sample, $e_{t}$, it is inserted into $\mathcal M$ and, if full, evicts the oldest entry (FIFO policy). A FIFO reservoir
$\mathcal M=\{e^{(n)}\}_{n=1}^{N_{\mathrm{buf}}}$
stores triples
\vspace{-2pt}
\[
e^{(n)}=
\bigl(
  \mathbf X^{(n)},    % input sequence
  \mathbf H^{(n)},    % ground-truth next channel
  \phi^{(n)}          % optional meta-data
\bigr),
\vspace{-2pt}
\]
where
$\mathbf H^{(n)}$ is the measured channel at
$t^{(n)}\!+\!\Delta$ and $\phi^{(n)}$ may encode
scenario ID, SNR, etc.
Each new sample
$\bigl(\mathbf X_{t},\mathbf H_{t+\Delta}\bigr)$
is added to $\mathcal M$ and, if full, the oldest entry is discarded.\\
\textbf{Interleaved Optimization. }For every incoming batch, $\mathcal D_{k}$, the composite
training set $\mathcal B = \mathcal D_{k}\cup\textsc{Sample}(\mathcal M,B_{\text{replay}})$,
where $B_{\text{replay}}$ replay items are drawn uniformly at random.
The predictor is updated by minimizing,
\begin{equation} \label{eq:er_loss}
\mathcal L_{\text{ER}} =
    \frac{1}{|\mathcal B|}
    \sum_{(\widehat{\mathbf H},\mathbf p_{\mathrm{tx}},\mathbf p_{\mathrm{rx}})\in\mathcal B}
    \!\!
    \bigl\|
      f_{\boldsymbol\Phi}
        (\widehat{\mathbf H})-
      \mathbf H_{\text{gt}}
    \bigr\|_{\mathrm F}^{2},
\end{equation}
thereby balancing plasticity to new fading statistics
against retention of earlier conditions.
In effect, ER allows the same predictor to track continuous
channel evolution, e.g.\ UMi-\textit{compact}
$\rightarrow$ UMi-\textit{dense}
$\rightarrow$ UMi-\textit{standard}, without catastrophic
forgetting, and to feed refined CSI back to the GRF estimator for the next cycle.\\
\textbf{Mini-batch Composition. }At global step $k$ the user equipment (UE) resides in cell $c(k)$ with local dataset $\mathcal D_{c(k)}$.
$\mathcal B_{\mathrm{curr}}\!\subset\!\mathcal D_{c(k)}$
and
$\mathcal B_{\mathrm{rep}}\!\subset\!\mathcal M$ is drawn
to form
$\mathcal B=\mathcal B_{\mathrm{curr}}\cup\mathcal B_{\mathrm{rep}}$.
Let $\bm{\theta}$ denote the current network parameters
($\bm{\theta}\!\equiv\!\boldsymbol\Phi$).
The overall loss is a convex mixture
\begin{equation}
\label{eq:er_total_loss}
\mathcal L_{\mathrm{tot}}(\bm{\theta})=
\lambda\,\mathcal L_{\mathrm{curr}}(\bm{\theta},\mathcal B_{\mathrm{curr}})
+(1-\lambda)\,
\mathcal L_{\mathrm{rep}} (\bm{\theta},\mathcal B_{\mathrm{rep}}),
\end{equation}
with mixing coefficient $\lambda\!\in\![0,1]$.
Both terms in~\eqref{eq:er_total_loss} use the
normalized mean-square error (NMSE):
\begin{align}
\mathcal L_{\mathrm{curr}} &=
\frac{1}{|\mathcal B_{\mathrm{curr}}|}
\sum_{(\mathbf X,\mathbf H)\in\mathcal B_{\mathrm{curr}}}
\frac{\bigl\|\mathbf H-f_{\bm{\theta}}(\mathbf X)\bigr\|_{F}^{2}}
     {\|\mathbf H\|_{F}^{2}}, \\
\mathcal L_{\mathrm{rep}} &=
\frac{1}{|\mathcal B_{\mathrm{rep}}|}
\sum_{(\mathbf X,\mathbf H)\in\mathcal B_{\mathrm{rep}}}
\frac{\bigl\|\mathbf H-f_{\bm{\theta}}(\mathbf X)\bigr\|_{F}^{2}}
     {\|\mathbf H\|_{F}^{2}}.
\end{align}

\textbf{6G Twin Channel Estimator. }During each prediction horizon, $(t,,t+\Delta)$ the ER–enhanced predictor delivers the look-ahead CSI $\widehat{\mathbf H}_{t+\Delta}$ to the resource scheduler, while its instantaneous error with the subsequent GRF update is back-propagated to refine both the predictor and the Gaussian field. This closed GRF–ER feedback loop (i) preserves high-fidelity CSI in the gaps between consecutive GRF renderings and (ii) tracks non-stationary propagation changes, e.g.\ UMi–\textit{compact}$\rightarrow$UMi–\textit{dense}, without catastrophic forgetting, thereby sustaining 6G-level accuracy throughout mobility and hand-over events.

\textbf{Reservoir Sampling. }As the UE traverses successive UMi–$\{\textit{compact},\textit{dense},\textit{standard}\}$ cells it generates a time–ordered stream  
$e_{t}=\bigl(\mathbf X_{t},\mathbf H_{t+\Delta}\bigr)$,  
where  
$\mathbf X_{t}\!\in\!\mathbb C^{2\times T\times N_{t}\times N_{r}}$ stacks the \emph{past} $T$ GRF estimates and  
$\mathbf H_{t+\Delta}\!\in\!\mathbb C^{2\times N_{t}\times N_{r}}$ is the \emph{next-slot} ground-truth channel. 
A replay reservoir $\mathcal M$ of fixed size $N_{\mathrm{buf}}$ (small enough to be cached at the serving gNB stores a uniform sample of all observations seen so far. 
Classical reservoir sampling guarantees
\[
\Pr\!\bigl[e_{i}\!\in\!\mathcal M_{t}\bigr]=\frac{N_{\mathrm{buf}}}{t},
\qquad i=1,\dots,t,
\]
So a channel recorded in the first cell is as likely to be rehearsed as one collected moments before the current GRF update. 

\emph{1) Data acquisition.}  
Upon arrival of $e_{t}$ the algorithm inserts it into $\mathcal M$ with probability $\tfrac{N_{\mathrm{buf}}}{t}$; otherwise the sample is discarded. 

\emph{2) Mini-batch assembly.}  
Each stochastic gradient descent (SGD) step forms 
$\mathcal B_{k}=\mathcal B_{\mathrm{curr}}\cup\mathcal B_{\mathrm{rep}}$,  
where  
$\mathcal B_{\mathrm{curr}}\!\subset\!\mathcal D^{(k)}$ is drawn from the live cell trace and  
$\mathcal B_{\mathrm{rep}}\!\subset\!\mathcal M$ from the reservoir. 
The predictor parameters $\boldsymbol\Phi$ are updated by minimizing the mixed NMSE of Eq.~\eqref{eq:er_loss}. 

\emph{3) Hand-over persistence.}  
When the UE hands off from cell $k$ to $k\!+\!1$, the same buffer $\mathcal M$ travels with it, ensuring that previously visited UMi statistics continue to influence training even though they are no longer observed. 
Coupled with the GRF estimator, this reservoir-based replay maintains a history-balanced memory, allowing the 6G-Twin channel estimator to track distribution shifts without sacrificing legacy performance.

\begin{algorithm}[t]
\caption{6G‐Twin Channel Estimation \& Prediction (GRF + ER/LARS)}
\label{alg:grf_er}
\begin{algorithmic}[1]
\Require buffer size $N_{\mathrm{buf}}$, mixing weight $\lambda$, 
         sampling mode $s\!\in\!\{\textsc{Uniform},\textsc{LARS}\}$,\\
         learning rates $\eta_{\mathrm{pred}}$, $\eta_{\mathrm{grf}}$,  
         GRF refresh period $\Delta_{\mathrm{GRF}}$
\State Initialize replay buffer $\mathcal M\!\leftarrow\!\varnothing$, counter $t\!\leftarrow\!0$
\State Initialize GRF parameters $\Theta_{\mathrm{GRF}}$, predictor parameters $\bm\Phi$
\Function{\textsc{Insert}}{$\mathbf X,\mathbf H,\ell$}
    \State $t \gets t+1$
    \If{$|\mathcal M|<N_{\mathrm{buf}}$}
        \State $\mathcal M\gets\mathcal M\cup\{(\mathbf X,\mathbf H,\ell)\}$
    \ElsIf{$\mathrm{rand}()<\tfrac{N_{\mathrm{buf}}}{t}$}       \Comment{reservoir test}
        \If{$s=\textsc{LARS}$}
            \State pick victim $v$ with prob.\! $\Pr[v=i]\propto(\ell_i+\varepsilon)^{-1}$  \label{line:lars}
        \Else
            \State $v\gets\mathrm{randint}(1,|\mathcal M|)$
        \EndIf
        \State $\mathcal M[v]\gets(\mathbf X,\mathbf H,\ell)$
    \EndIf
\EndFunction
\For{each visited cell $k$}
    \For{each measurement $(\mathbf p_{\mathrm{tx}},\mathbf p_{\mathrm{rx}},\mathbf H_{t})$}
        \If{$t\;\bmod\;\Delta_{\mathrm{GRF}}=0$}               \Comment{full GRF update}
            \State $\widehat{\mathbf H}_{t}\gets
                   \textsc{RenderGRF}(\Theta_{\mathrm{GRF}},
                                      \mathbf p_{\mathrm{rx}},
                                      \mathbf p_{\mathrm{tx}})$
            \State $\Theta_{\mathrm{GRF}}\gets
                   \Theta_{\mathrm{GRF}}
                   -\eta_{\mathrm{grf}}\,
                   \nabla_{\Theta_{\mathrm{GRF}}}
                   \|\widehat{\mathbf H}_{t}-\mathbf H_{t}\|_{F}^{2}$
        \EndIf
        \State Form window
               $\mathbf X_{t}=
                 [\widehat{\mathbf H}_{t-T+1},\dots,\widehat{\mathbf H}_{t}]$
        \State $\tilde{\mathbf H}_{t+\Delta}\gets
               f_{\bm\Phi}(\mathbf X_{t})$                         \Comment{look-ahead CSI}
        \State $\ell_{t}\gets
               \|\mathbf H_{t+\Delta}-\tilde{\mathbf H}_{t+\Delta}\|_{F}^{2}\big/
               \|\mathbf H_{t+\Delta}\|_{F}^{2}$                    \Comment{NMSE}
        \State \textsc{Insert}($\mathbf X_{t},\mathbf H_{t+\Delta},\ell_{t}$)
        \If{ready to update predictor}
            \State Sample $\mathcal B_{\mathrm{curr}}\!\subset\!\mathcal D^{(k)}$,
                   $\mathcal B_{\mathrm{rep}}\!\subset\!\mathcal M$
            \State Compute mixed loss
                   $\mathcal L_{\mathrm{tot}}$ via Eq.~\eqref{eq:er_loss}
            \State $\bm\Phi\gets
                   \bm\Phi-\eta_{\mathrm{pred}}\,
                   \nabla_{\bm\Phi}\mathcal L_{\mathrm{tot}}$
        \EndIf
    \EndFor
\EndFor
\end{algorithmic}
\end{algorithm}

\textbf{Loss Aware Reservoir Sampling (LARS). }Uniform reservoir sampling treats every past observation equally; however, deep fades or cell-edge links precisely those that limit throughput may then be under-represented. Therefore, uniform eviction is replaced with LARS~\cite{mall2023change,kumari2022retrospective}. 
Immediately after the forward pass on mini-batch, $\mathcal B_{k}$ per-sample NMSE is recorded  
$\ell_{i}=\|\mathbf H_{i}-f_{\bm\Phi}(\mathbf X_{i})\|_{F}^{2}\big/\|\mathbf H_{i}\|_{F}^{2}$  
for each element $e_{i}\!\in\!\mathcal B_{k}$. 
If the buffer is full ($|\mathcal M|=N_{\mathrm{buf}}$) and a new sample $e_{t}$ is \emph{selected} for insertion with reservoir probability $N_{\mathrm{buf}}/t$, LARS evicts an existing entry $v$ drawn according to  
\begin{equation}
\Pr[v=i]=
{(\,\ell_{i}+\varepsilon)^{-1}} /
     {\displaystyle\sum_{j=1}^{N_{\mathrm{buf}}}
       (\ell_{j}+\varepsilon)^{-1}},
\label{eq:lars_prob}
\end{equation}
with $\varepsilon>0$ preventing division by zero. 
Hence, items whose prediction loss is already small are more likely to be removed, while hard-to-predict channels (large loss value $\ell_{i}$) persist longer. This bias keeps rare yet performance-critical propagation states deep fades, rich scattering, extreme cell-edge SNRs, actively rehearsed, enhancing the robustness of the GRF-aware continual predictor.

\section{Non-Linear Precoder}
At scheduler time $t$, the GRF estimator delivers the most recent CSI snapshot $\widehat{\mathbf H}_{u,n,t}\in\mathbb C^{L_y\times L_{x,u}}$ and the CL predictor provides a $\Delta$-step look-ahead $\widehat{\mathbf H}_{u,n,t+\Delta}$ together with an error summary (e.g., innovation covariance) $\mathbf\Sigma_{u,n,t+\Delta}$. The true channel is modeled as
\begin{equation}
\mathbf H_{u,n,t+\Delta}=\widehat{\mathbf H}_{u,n,t+\Delta}+\mathbf E_{u,n,t+\Delta}
\end{equation}

\begin{equation}
\mathbb E[\operatorname{vec}(\mathbf E)\operatorname{vec}(\mathbf E)^{\mathrm H}] = \mathbf\Sigma_{u,n,t+\Delta},
\label{eq:pred_model}
\end{equation}
and use either (i) a worst-case ellipsoidal set $\{\mathbf E:\ \|\mathbf\Sigma^{-1/2}\operatorname{vec}(\mathbf E)\|_2\le \rho\}$, or (ii) a chance-constraint with violation level $\varepsilon$.

\subsection{Energy–Optimal Nonlinear Precoding on the MAC Polymatroid}
For each user $u$ and tone $n$ let $\mathbf R_{xx}(u,n)\succeq\mathbf 0$ denote the (possibly multi-stream) transmit covariance. Denote noise covariance by $\mathbf R_{nn}=\sigma^2\mathbf I_{L_y}$. The achievable \emph{per-tone} rate tuple under successive interference cancellation (SIC) must lie in the MAC capacity region, which is a polymatroid: for every subset $\mathcal T\subseteq\mathcal U$,
\begin{equation}
\begin{split}
\sum_{u\in\mathcal T} b_{u,n} \ \le\ &
\log_2\!\left|\mathbf R_{nn}
+\sum_{u\in\mathcal T}\widehat{\mathbf H}_{u,n}\mathbf R_{xx}(u,n)\widehat{\mathbf H}_{u,n}^{\mathrm H}\right| \\
&\ - \log_2\!\left|\mathbf R_{nn}\right|.
\end{split}
\label{eq:mac_poly}
\end{equation}
Summing over tones yields the standard MAC polymatroid for the block. This description is exact for Gaussian codebooks and SIC and is independent of the decoding order.
Weighted sum of user energies is minimized while satisfying minimum rate targets $\mathbf b_{\min}\succeq\mathbf 0$:
\begin{equation}
\label{eq:energy_min_det}
\begin{aligned}
\min_{\{\mathbf R_{xx}(u,n)\succeq\mathbf 0\}} \quad
& \sum_{u=1}^U\sum_{n=1}^N w_u\,\operatorname{tr}\!\big(\mathbf R_{xx}(u,n)\big) \\
\text{s.t.}\quad
& \mathbf b=\sum_{n=1}^{N}[b_{1,n},\dots,b_{U,n}]^{\!\top}\ \succeq\ \mathbf b_{\min},\\
& \text{(\ref{eq:mac_poly}) holds for all } \mathcal T\subseteq\mathcal U,\ \forall n\in\mathcal N.
\end{aligned}
\end{equation}
Problem \eqref{eq:energy_min_det} is a convex semidefinite program (SDP): the objective is linear in $\{\mathbf R_{xx}\}$, constraints are concave in the covariances by MIMO-MAC capacity, and PSD constraints are conic. The formulation is order-free; SIC ordering is recovered from the dual variables.\\

\textbf{Dual structure and SIC order.}
Associate multipliers $\{\theta_u\ge 0\}$ to the per-user minimum-rate constraints and $\{\lambda_{\mathcal T,n}\ge 0\}$ to \eqref{eq:mac_poly}. At optimality, only a \emph{chain} of subset constraints $\emptyset\subset\mathcal T_1\subset\cdots\subset\mathcal T_U=\mathcal U$ is active on each tone, yielding a corner point of the polymatroid that corresponds to a specific SIC order $\boldsymbol\pi$; the active chain determines $\boldsymbol\pi$ and vice-versa. Practically, the globally optimal $\boldsymbol\pi$ can be inferred by sorting the optimal duals $\{\theta_u^\star\}$ (ties broken by the active-set structure). This is the standard order–duality relationship for (weighted) MAC boundary points and underpins efficient order recovery without combinatorial search.

\subsection{Recovering Decoding Vectors \texorpdfstring{$\boldsymbol\theta_{u,n}$}{theta} and Implementing Nonlinear Precoding} \label{subsec:decoding-order}
While \eqref{eq:energy_min_det} determines optimal covariances $\{\mathbf R_{xx}(u,n)\}$ and the SIC order $\boldsymbol\pi$, practical receivers need \emph{decoding vectors} for each stream. Suppose user $u$ sends $d_{u,n}$ streams on tone $n$ via the rank-$d_{u,n}$ factorization
\begin{equation}
\mathbf R_{xx}(u,n)=\sum_{s=1}^{d_{u,n}} p_{u,n,s}\,\mathbf f_{u,n,s}\mathbf f_{u,n,s}^{\mathrm H},\quad \|\mathbf f_{u,n,s}\|_2=1,\ p_{u,n,s}\ge 0.
\end{equation}
Under the optimal SIC order $\boldsymbol\pi$, let $\mathcal I_{n}^{(k)}$ be the set of users not yet decoded before stage $k$ (i.e., residual interferers when decoding user $\pi_k$). The linear MMSE-SIC decoder for stream $(u\!=\!\pi_k,s)$ is
\begin{equation}
\boldsymbol\theta_{u,n,s}
\ \propto\
\Big(\mathbf R_{nn}+\textstyle\sum_{i\in \mathcal I_{n}^{(k)}}\widehat{\mathbf H}_{i,n}\mathbf R_{xx}(i,n)\widehat{\mathbf H}_{i,n}^{\mathrm H}\Big)^{-1}
\widehat{\mathbf H}_{u,n}\,\mathbf f_{u,n,s}.
\label{eq:mmse_sic}
\end{equation}
Equations \eqref{eq:energy_min_det}–\eqref{eq:mmse_sic} fully specify the transmit covariances, the optimal SIC sequence, and the per-stream decoding vectors. When the downlink (BC) is ultimately to be served, the MAC solution is mapped to the BC via uplink–downlink duality and realize the non-linear transmitter with \emph{dirty-paper coding} (DPC) or its practical approximations—Tomlinson–Harashima precoding (THP) and vector perturbation (VP). DPC achieves the BC capacity region; THP/VP are near-DPC at substantially lower complexity~\cite{Cioffi_GDFE_Ch5, Cioffi1997}.\\
\textbf{Worst-case (ellipsoidal) robust Semi-Definite Programming (SDP).}
Replace $\widehat{\mathbf H}_{u,n}$ in \eqref{eq:mac_poly} by $\widehat{\mathbf H}_{u,n}+\mathbf E_{u,n}$ and require the inequality to hold for all $\mathbf E_{u,n}$ in the ellipsoid
$\|\mathbf\Sigma^{-1/2}\operatorname{vec}(\mathbf E_{u,n})\|_2\le \rho$.
Using the S-procedure, each set-membership log-det constraint admits a safe LMI relaxation, producing a tractable SDP; for linear-beamforming special cases, semidefinite relaxation (SDR) is often tight.\\
\textbf{Chance-constrained design.}
Enforce, for each $\mathcal T,n$,
\begin{equation}
\label{eq:chance}
\begin{split}
\Pr\!\Bigg\{\sum_{u\in\mathcal T} b_{u,n}\ \le\
\log_2\Big|\mathbf R_{nn}+\!\sum_{u\in\mathcal T}(\widehat{\mathbf H}_{u,n}+\mathbf E_{u,n})\mathbf R_{xx}(u,n)\\
\cdot(\widehat{\mathbf H}_{u,n}+\mathbf E_{u,n})^{\mathrm H}\Big|
-\log_2|\mathbf R_{nn}|\Bigg\}\ \ge\ 1-\varepsilon .
\end{split}
\end{equation}
Directly optimizing \eqref{eq:chance} is intractable; \emph{safe convex approximations} are adopted based on measure-concentration inequalities or scenario/sample approximations, which lead to explicit LMIs or conic constraints and preserve feasibility with probability at least $1-\varepsilon$. The resulting program remains convex in $\{\mathbf R_{xx}\}$.

\begin{figure}[t!]
    \centering
    \includegraphics[width=\linewidth]{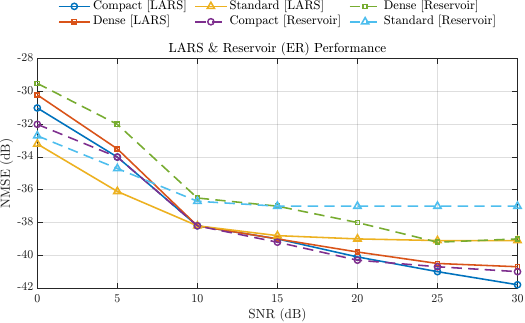}
    \caption{SNR [dB] vs. NMSE [dB] for Experience Replay.}
    \label{fig:continual_cp}
\end{figure}
\section{Complexity Analysis}
\noindent\textbf{GRF Channel Estimation.}
Let $N_G$ be the number of Gaussian primitives and $(N_t,N_r)$ the Tx/Rx antenna counts used in GRF rendering (Eq.~\eqref{eq:grf_render}). The rendering cost is
\begin{equation}
\mathcal C_{\text{GRF-render}}=\Theta\!\big(N_G\,N_t\,N_r\big).
\end{equation}
With GRF refresh period $\Delta_{\text{GRF}}$ (Alg.~\ref{alg:grf_er}), the amortized per-slot cost is
\begin{equation}
\mathcal C_{\text{GRF,slot}}=\Theta\!\big(\tfrac{1}{\Delta_{\text{GRF}}}\,N_G\,N_t\,N_r\big).
\end{equation}

\noindent\textbf{Continual Channel Prediction.}
Let $B=|\mathcal B|$ be the mixed mini-batch size, $T$ the window length, and $N_{\text{buf}}$ the replay-buffer size (Sec.~\textit{Reservoir Sampling}). Denote the model-dependent forward/backward cost by $\mathcal C_{\text{pred}}(B;T,N_t,N_r)$. One CL update step incurs
\begin{equation}
\mathcal C_{\text{CL-step}}
=\mathcal C_{\text{pred}}(B;T,N_t,N_r)
+\Theta(B)+\Theta(1)+\Theta(N_{\text{buf}}),
\end{equation}
where $\Theta(B)$ is NMSE evaluation over $\mathcal B$, $\Theta(1)$ is the reservoir test, and $\Theta(N_{\text{buf}})$ is the LARS eviction (Eq.~\eqref{eq:lars_prob}).

\noindent\textbf{Non-Linear Precoder (energy–sum minimization on MAC polymatroid).}
For each tone $n$, constraint \eqref{eq:mac_poly} requires a log-det of an $L_y\times L_y$ matrix. Naively checking all subset constraints $(2^U\!-\!1)$ per tone gives
\begin{equation}
\mathcal C_{\text{poly-check}}=\Theta\!\big(N\,(2^U-1)\,L_y^{3}\big).
\end{equation}
The SIC order recovery from duals (sorting $\{\theta_u^\star\}$) costs
\begin{equation}
\mathcal C_{\text{order}}=\Theta\!\big(U\log U\big).
\end{equation}
MMSE–SIC decoding vectors (Eq.~\eqref{eq:mmse_sic}) require one $L_y\times L_y$ solve per decoded user per tone:
\begin{equation}
\mathcal C_{\text{MMSE-SIC}}=\Theta\!\big(N\,U\,L_y^{3}\big).
\end{equation}

\noindent\textbf{One closed-loop adaptation cycle.}
With GRF amortized by $\Delta_{\text{GRF}}$, a full slot that includes prediction, optimization, order recovery, and decoders has
\begin{equation}
\begin{aligned}
\mathcal C_{\text{cycle}}
&=\Theta\!\big(\tfrac{1}{\Delta_{\text{GRF}}}\,N_G\,N_t\,N_r\big)
+\mathcal C_{\text{pred}}(B;T,N_t,N_r)
+\\\Theta\!\big(B+N_{\text{buf}}) 
&\quad+\Theta\!\big(N\,(2^U-1)\,L_y^{3}\big)
+\Theta\!\big(N\,U\,L_y^{3}+U\log U\big).
\end{aligned}
\end{equation}

\begin{figure}
    \centering
    \includegraphics[width=\linewidth]{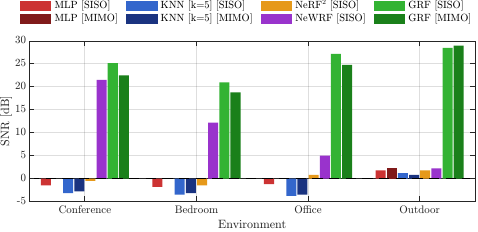}
    \caption{Comparison of SNR [dB] across environments and methods. Our proposed \textbf{GRF} achieves consistently higher SNR in both SISO and MIMO modes, outperforming all prior models across diverse environments.}
    \label{fig:performance_comparison}
\end{figure}
\section{Quantitative Result Analysis}
\subsection{GRF Channel Estimation}
All methods are evaluated using the signal–to–noise ratio (SNR), defined as
\begin{equation}
    \text{SNR}_{\mathrm{H,dB}}
    = 10\log_{10}\frac{\lVert\mathbf{H}_{\text{gt}}\rVert_F^{2}}
                               {\lVert\mathbf{H}_{\text{pred}}-\mathbf{H}_{\text{gt}}\rVert_F^{2}},
    \label{eq:snr}
\end{equation}
where $\mathbf{H}_{\text{gt}}$ and $\mathbf{H}_{\text{pred}}$ are the ground‑truth and predicted channel‑gain tensors, respectively. A higher value indicates more accurate channel reconstruction. GRF is compared with four strong alternatives:  
NeWRF\,\cite{lu2024newrf},  
NeRF$^{2}$\,\cite{Zhao_2023},  
a vanilla basleine MLP,  
and a $k$‑nearest‑neighbours (KNN) estimator. 
Each model is tested in both SISO and MIMO configurations. As summarised in Fig.\,\ref{fig:performance_comparison}, GRF delivers the highest SNR in every indoor setting. 
In SISO mode it attains \textbf{25.23\,dB}, \textbf{21.14\,dB}, and \textbf{26.53\,dB} in the \emph{conference room}, \emph{bedroom}, and \emph{office}, respectively—an average gain of \underline{10.9\,dB} over the next best approach (NeWRF). 
Even in the more demanding MIMO case, GRF maintains a clear margin, achieving \textbf{22.73\,dB}, \textbf{19.60\,dB}, and \textbf{24.78\,dB} in the same environments. 
Traditional learning‑based baselines, particularly MLP and KNN, fall below $0$\,dB in most indoor scenes, confirming that simple interpolators lack the inductive bias required for fine‑grained channel reconstruction.

The outdoor scenario further highlights GRF’s scalability. 
It achieves \textbf{28.32\,dB} (SISO) and \textbf{27.92\,dB} (MIMO), which is \textasciitilde26\,dB higher than NeWRF (2.03\,dB) and \textasciitilde27\,dB above NeRF$^{2}$ (1.40\,dB). 
These results indicate that GRF can effectively model both rich multipath indoor channels and the sparser, large‑scale characteristics of outdoor propagation. Across all eight test cases, GRF is consistently the top performer, underscoring the value of integrating Gaussian radio fields with neural implicit representations for high‑fidelity channel prediction. The substantial margin—often exceeding an order of magnitude in MSE—demonstrates that GRF generalizes better than prior NeRF‑style methods and conventional learning‑based baselines, particularly in challenging MIMO and large‑scale settings.

\begin{figure*}[t!]
  \centering
  % First row
  \begin{subfigure}[t]{0.32\textwidth}
      \centering
      \includegraphics[width=\textwidth]{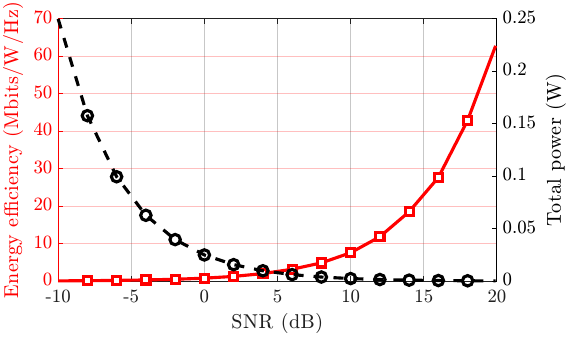}
      \caption{Energy efficiency and decreasing total transmit power vs. SNR for indoor scenarios.}
      \label{fig:subfig1}
  \end{subfigure}%
  \hspace{0.5em}
  \begin{subfigure}[t]{0.32\textwidth}
      \centering
      \includegraphics[width=\textwidth]{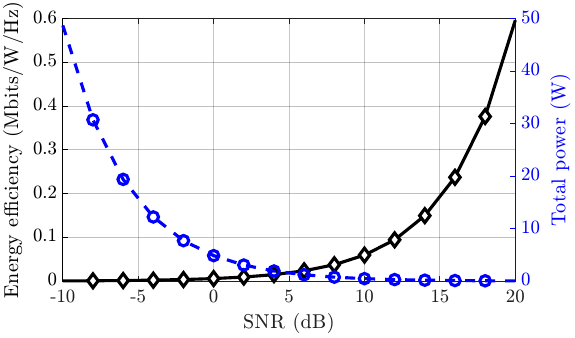}
      \caption{Energy efficiency and decreasing total transmit power vs. SNR for Uma scenarios.}
      \label{fig:subfig2}
  \end{subfigure}%
   \hspace{0.5em}
  \begin{subfigure}[t]{0.32\textwidth}
      \centering
      \includegraphics[width=\textwidth]{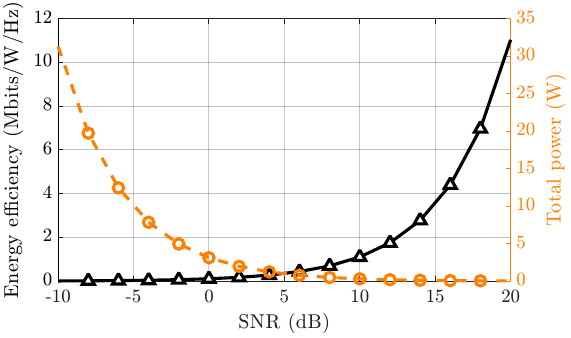}
      \caption{Energy efficiency and decreasing total transmit power vs. SNR for Umi scenarios.}
      \label{fig:subfig2}
  \end{subfigure}%

  \caption{Energy efficacy vs. SNR(db) and Total Transmit Power (W) vs. SNR for various scenarios for the \texttt{minPMAC} non linear precoder design.}

  \label{fig_8}
\end{figure*}
\subsection{Continual Adaptation with Experience Replay}
Figure \ref{fig:continual_cp} reports NMSE versus SNR when the predictors are trained in a continual‑learning loop that interleaves current samples with a bounded replay buffer. Two buffer management strategies are compared: LARS (loss‑aware reservoir sampling; solid lines), which preferentially retains high‑loss/hard channel states, and uniform Reservoir replay (dashed lines). Unlike the baseline models in Fig. \ref{fig:baseline_cp}, where Standard and Compact models saturated near $-22\,$dB to $-25\,$dB and even Dense variants stalled around $-34\,$dB at high SNR, the online CL models rapidly adapt as conditions improve. By $\sim10\,$dB SNR all scenarios collapse to a low‑error band around $-38\,$dB, eliminating the baseline scenario-dependent spread.

In these experiments, uniform reservoir replay already improves over the static baseline: for example, the Standard model’s NMSE drops from approximately \(-25\,\mathrm{dB}\) when frozen to around \(-37\,\mathrm{dB}\) at high SNR. However, it plateaus there because random sampling fails to emphasize the rare, challenging fades. This gap highlights the extra value of LARS’s loss‑aware sampling, which keeps the most challenging fades. By continuously retraining and focusing on these hard examples, the predictor adapts to changing channel conditions and maintains low NMSE across all scenarios, even under high‑SNR conditions. LARS is adopted, which keeps the highest‑loss examples in the buffer. With LARS, models drive NMSE even lower: at \(30\,\mathrm{dB}\) SNR, Standard‑LARS reaches about \(-39\,\mathrm{dB}\), Dense‑LARS about \(-41\,\mathrm{dB}\), and Compact‑LARS about \(-42\,\mathrm{dB}\). These extra \(2\)–\(5\,\mathrm{dB}\) improvements beyond uniform replay highlight how loss‑aware sampling further adapts the predictor to clean channels and maintains low NMSE across all scenarios.

\subsection{Non-Linear Precoder Design}
Across all three environments (indoor, UMa, UMi), Fig~\ref{fig_8}'s two--axis plots place \emph{energy efficiency} (left $y$--axis, in Mbit/J/Hz) and \emph{total transmit power} (right $y$--axis, in W) against the operating SNR (in dB). As SNR increases, the \emph{minimum} power needed to satisfy the per--user rate vector $\mathbf b_{\min}$ \emph{decreases} monotonically, while the achieved energy efficiency \emph{increases} monotonically over the same SNR range.

In the indoor panel, energy efficiency rises steeply across the SNR sweep, entering the \emph{60–70 Mb/J/Hz} band by \(\mathrm{SNR}\approx 20\,\mathrm{dB}\); over the same range, the total transmit power required to meet the target minimum rates collapses into the \emph{20–50 mW} band. The curve transitions from near-zero efficiency at deep negative SNRs to roughly \(\mathbf{65\pm5~Mb/J/Hz}\) at \(20\,\mathrm{dB}\), while the total power falls from the \(\mathbf{0.20\text{–}0.25~W}\) bracket at \(-10\,\mathrm{dB}\) to about \(\mathbf{0.02\text{–}0.05~W}\) at \(20\,\mathrm{dB}\). This corresponds to reduction in required power over the plotted operating window, consistent with \texttt{minPMAC}’s ability to re-allocate energy and pick the globally optimal SIC order as the feasible MAC region expands with SNR.

In the urban-macro (UMa) plot, absolute scales are harsher, yet the qualitative behavior remains: energy efficiency climbs from near zero to approximately \(\mathbf{0.5\text{–}0.6~Mb/J/Hz}\) by \(20\,\mathrm{dB}\), while the total power to satisfy the same rate targets drops from tens of Watts at low SNR to the \(\mathbf{10\pm2~W}\) band at the high-SNR end. 

For UMi, the energy-efficiency trace increases into the \(\mathbf{10\text{–}12~Mb/J/Hz}\) range by \(20\,\mathrm{dB}\), while the total-power curve descends from the \(\mathbf{30\pm5~W}\) region at \(-10\,\mathrm{dB}\) to roughly \(\mathbf{5\text{–}8~W}\) at \(20\,\mathrm{dB}\). These readings imply an \(\mathbf{>\!4}\times\) reduction in required power along with a near \(\mathbf{10}\times\) rise in bits/J/Hz across the plotted SNRs. Compared with UMa, UMi achieves noticeably higher energy efficiency at the top end (owing to shorter ranges and improved spatial conditioning), while still delivering multi-Watt savings in total power when operated in the moderate-to-high SNR regime.

Across indoor, UMa, and UMi, the two-axis figures consistently show \emph{monotonically decreasing} total power and \emph{monotonically increasing} energy efficiency as SNR grows. Quantitatively, indoor achieves the largest absolute efficiency (peaking near \(\sim\!65~\mathrm{Mb/J/Hz}\) and reducing to a few tens of milliwatts), UMi lands in the \(\sim\!10\text{–}12~\mathrm{Mb/J/Hz}\) band with \(\sim\!5\text{–}8~\mathrm{W}\) at high SNR, and UMa reaches \(\sim\!0.5\text{–}0.6~\mathrm{Mb/J/Hz}\) with \(\sim\!8\text{–}12~\mathrm{W}\). These gains are attributable to \texttt{minPMAC}’s order-free convex allocation (energy vectors + subcarrier assignment) followed by dual-driven SIC ordering, which together convert the SNR expansion of the MAC region into lower Joules per delivered bit and, consequently, lower required transmit power at the same service targets.

\section{Conclusion and Future Work}\label{sec:conclusion}
Our experiments yield three core lessons that go beyond architecture minutiae: (i) GRF succeeds not because it is larger, but because a sparse, additive field of anisotropic Gaussians respects superposition, collapsing CSI synthesis to a single weighted sum and enabling sub-ms operation; (ii) continual prediction improves most when memory prioritizes hard states (deep fades, handovers), with LARS outperforming uniform replay by keeping failure modes in the buffer; and (iii) \texttt{minPMAC} decouples SIC order from covariance allocation and recovers the order from duals, delivering consistent energy savings and monotonic bits/J with SNR without combinatorial search. The most important empirical finding is multiplicative: GRF reduces pilot burden and stabilizes predictor targets; the predictor supplies look-ahead CSI that the scheduler can trust; and the scheduler’s energy savings feed back into cheaper, more frequent re-estimation, yielding better than additive gains. Operationally, three knobs consistently drove the quality–latency trade: 1) the GRF refresh cadence $\Delta_{\mathrm{GRF}}$ (balancing drift versus overhead), 2) the replay budget and its loss-aware policy (small buffers work if they retain hard samples), 3) and the minimum-rate vector $\mathbf b_{\min}$ (which selects the active face of the MAC polymatroid and thus power scaling). Problems arise when engineers ignore how frequency affects signals or overlook near-field effects. Systems also struggle when individual signal processing becomes too complex for large networks. The key insights include that 6G designers must understand physics, adapt to changing conditions, and use stable mathematical structures. Our combined GRF, CL(LARS), and minPMAC approach puts this principle into practice.

Future work will (a) extend GRF to multi-cell cooperative twins that share compact summaries to improve edge throughput and handover stability, (b) implement uncertainty-aware, cross-time/frequency pilot placement to further cut overhead and speed re-acquisition after abrupt changes, (c) add wideband and near-field support via frequency-dependent primitives and spherical-wave rendering, (d) integrate robust and chance-constrained scheduling using predictor innovation covariances for reliability guarantees, and (e) co-design CUDA/ASIC kernels for GRF, replay, and log-det blocks to ensure sub-ms closed-loop operation on realistic gNB budgets.

\bibliographystyle{IEEEtran}
\bibliography{main}
\end{document}